\documentclass[aps,twocolumn,showpacs]{revtex4}
\begin{document}
\title{BBGKY Hierarchy Underlying Many Particle Quantum Mechanics}
\author{G. Kaniadakis}
\email{giorgio.kaniadakis@polito.it} \affiliation{Dipartimento di
Fisica and Istituto Nazionale di
 Fisica della Materia, \\ Politecnico di Torino,
Corso Duca degli Abruzzi 24, 10129 Torino, Italy}
\date{\today}

\begin{abstract}
Recently, the one particle quantum mechanics has been obtained in
the framework of an entirely classical subquantum kinetics. In the
present paper we argue that, within the same scheme and without
any additional assumption, it is possible to obtain also the
$n$-particle non relativistic quantum mechanics. The main goal of
the present effort is to show that the classical BBGKY
hierarchical equation, for the $n$-particle reduced distribution
function, is the ancestor of the $n$-particle Schr\"odinger
equation. On the other hand we show that within the scenario of
the subquantum structure of quantum particle, the Fisher
information measure emerges naturally in quantum mechanics.

\end{abstract}
 \maketitle

In ref. \cite{GK}, it is shown that quantum mechanics can be
obtained in a self consistent scheme of an entirely classical
many body physics. Within this framework the mean features of
quantum mechanics (i.e. the probabilistic nature of the quantum
description, the quantum potential, the Schr\"odinger equation,
the quantum operators, the Heisenberg uncertainty principle) were
obtained starting from a classical subquantum kinetics in the
phase space and without invoking any additional principle.
Furthermore the fundamental constant $\hbar$ emerges naturally as
an integration constant and represents a free parameter for the
theory. Within this theory, the astonishing scenario of a
subquantum structure for the quantum particle emerges
spontaneously. The quantum particle turns out to have an internal
structure and a spatial dispersion and appears to be composed by
$N$ identical point like and interacting subquantum objects, the
monads, obeying to the laws of Newtonian physics. The theory can
be viewed as describing a mechanism which permits to construct
the quantum particle starting from its constituents and then
quantum mechanics appears to describe a physical reality.

The statistical ensemble of these $N$ monads is described in the
phase space through the distribution function $f=f(t,
\mbox{\boldmath $x$} , \mbox{\boldmath $v$})$ whose time
evolution is  governed by the kinetic equation ${\cal
L}\,f=C(f)$, being ${\cal L}$ the Liouville operator and $C(f)$
the collision integral that takes into account the intermonad
interactions. We don't make any assumption about the structure of
the collision integral and the nature of the interaction between
the monads except that, during the point-like collisions the
monad number, momentum and energy are conserved. The projection
of the above phase space kinetics into the physical space
produces a hydrodynamics which naturally leads to the one
particle spinless quantum mechanics when the external force field
is conservative.

Recently in ref. \cite{GK2}, it is shown that the above kinetic
approach can be easily adopted to treat the quantum particle with
spin in an external electromagnetic field. Of course in this way
the spinning quantum mechanics can be obtained.

The energy conservation during the collisions considered in the
present scheme plays a very important role and could be also taken
into account in the stochastic models of  quantum mechanics as it
has been recently observed in ref. \cite{CG}.

A question which naturally arises  at this point is if it is
possible to extend the kinetic formalism of ref. \cite{GK} in
order to treat non relativistic many particle quantum systems.
This point is neglected in the literature where  almost the
totality of works devoted to the derivation and interpretation of
quantum mechanics concern the single particle.

The main goal of the present paper is to show that it is possible
to obtain, within the theory developed in ref. \cite{GK} and
without any additional assumption,  also the $n$-particle quantum
mechanics starting from the classical kinetics governed in the
phase space by the BBGKY hierarchy.

{\it BBGKY Hierarchy:} Let us consider a statistical ensemble
constituted by ${\cal N}$ identical and indistinguished
point-like particles (monads) obeying the laws of classical
physics. It is well known that the $n$-particle reduced
distribution function $f_n=f_n(t,\mbox{\boldmath
$x$}_1,\mbox{\boldmath $v$}_1, \mbox{\boldmath
$x$}_2,\mbox{\boldmath $v$}_2,...,\mbox{\boldmath
$x$}_n,\mbox{\boldmath $v$}_n)$ obeys the following BBGKY
hierarchical evolution equation \cite{BA}
\begin{eqnarray}
\frac{\partial {f_n}}{\partial t}
 + \sum_{i=1}^n \mbox{\boldmath $v$}_{i}\,
\frac{\partial {f_n}}{\partial \mbox{\boldmath $x$}_{i}} +
\sum_{i=1}^n \frac{\mbox{\boldmath $F$}_{i}}{\mu}
\,\frac{\partial {f_n}}{\partial \mbox{\boldmath $v$}_{i}}
=\sum_{i=1}^n C_i(f_{n+1}) \ \ , \label{1}
\end{eqnarray}
being $\mbox{\boldmath $F$}_i=-\partial {{\cal V}_i}/\partial
\mbox{\boldmath $x$}_i$ and
\begin{eqnarray}
{\cal V}_{i}={\cal V}^{(ex)}(\mbox{\boldmath $x$}_{i})+
\frac{1}{2}\sum_{j=1 \atop j\neq i}^n {\cal
V}^{(in)}(|\mbox{\boldmath $x$}_{i}-\mbox{\boldmath $x$}_{j}|)\ \
. \label{2}
\end{eqnarray}
$C_i(f_{n+1})$ are the collision integrals which can be expressed
in term of the function $f_{n+1}$. The assumption that during the
point-like collisions the monad number, momentum and energy are
conserved,  implies that the three functions $g_1(\mbox{\boldmath
$v$}_i)=1$, $g_2(\mbox{\boldmath $v$}_i)=\mbox{\boldmath $v$}_i$
and $g_3(\mbox{\boldmath $v$}_i)=\mbox{\boldmath $v$}_i^2$ are
the collision invariants of $C_i(f_{n+1})$
\begin{equation}
\int g_k(\mbox{\boldmath $v$}_i)\,C_i(f_{n+1})\, d^{3} v_i=0  \ \
; \ \ k=1,2,3  \ \ . \label{3}
\end{equation}

Let us consider the $6n$-dimensional phase space (\mbox{\boldmath
$x$},\mbox{\boldmath $v$}) being $\mbox{\boldmath
$x$}=(\mbox{\boldmath $x$}_1,\mbox{\boldmath
$x$}_2,...,\mbox{\boldmath $x$}_n)$ and $\mbox{\boldmath
$v$}=(\mbox{\boldmath $v$}_1,\mbox{\boldmath
$v$}_2,...,\mbox{\boldmath $v$}_n)$. We introduce the two
gradient operators $\partial/\partial\mbox{\boldmath
$x$}=(\partial/\partial\mbox{\boldmath
$x$}_1,\partial/\partial\mbox{\boldmath
$x$}_2,...,\partial/\partial\mbox{\boldmath $x$}_n)$ and
$\partial/\partial\mbox{\boldmath
$v$}=(\partial/\partial\mbox{\boldmath
$v$}_1,\partial/\partial\mbox{\boldmath
$v$}_2,...,\partial/\partial\mbox{\boldmath $v$}_n)$. Then we
define the $3n$-dimensional force $\mbox{\boldmath
$F$}=(\mbox{\boldmath $F$}_1,\mbox{\boldmath
$F$}_2,...,\mbox{\boldmath $F$}_n)$ which can be derived from a
potential
\begin{eqnarray}
\mbox{\boldmath $F$}=-\frac{\partial
 \,{\cal V}}{\partial{\mbox{\boldmath $x$}}}
 \ \  ; \ \  {\cal
V}=\sum_{i=1}^n {\cal V}_{i} \ \ , \label{4}
\end{eqnarray}

The BBGKY hierarchical equation (\ref{1}), governing the
evolution of the $n$-particle reduced distribution function
$f_n=f_n(t,\mbox{\boldmath $x$},\mbox{\boldmath $v$})$, can be
written in the following compact form
\begin{equation}
\frac{\partial {f_n}}{\partial t}
 + \mbox{\boldmath $v$}\cdot
\frac{\partial {f_n}}{\partial \mbox{\boldmath $x$}} +
\frac{\mbox{\boldmath $F$}}{\mu} \cdot \frac{\partial
{f_n}}{\partial \mbox{\boldmath $v$}} = C(f_{n+1}) \ \ . \label{5}
\end{equation}
The collision integral
\begin{equation} C(f_{n+1})=\sum_{i=1}^n
C_i(f_{n+1}) \ \ , \label{6}
\end{equation}
admits the three collision invariants $g_1(\mbox{\boldmath
$v$})=1$, $g_2(\mbox{\boldmath $v$})=\mbox{\boldmath $v$}$ and
$g_3(\mbox{\boldmath $v$})=\mbox{\boldmath $v$}^2$ satisfying the
conditions
\begin{equation}
\int g_k(\mbox{\boldmath $v$})\,C(f_{n+1})\, d^{3n} v=0  \ \ ; \ \
k=1,2,3  \ \ . \label{7}
\end{equation}
as one can verify immediately starting from Eq. (\ref{3}).

{\it Hydrodynamics:} We consider now the projection of the
kinetics from the $6n$-dimensional phase space into the
$3n$-dimensional space of the coordinates $\mbox{\boldmath $x$}$,
where the system is described through the distribution function
$\rho_{_{\scriptstyle n}}(t, \mbox{\boldmath $x$})= \int f_n(t,
\mbox{\boldmath $x$} , \mbox{\boldmath $v$})\, d^{3n} v $.  After
recalling that the total value of a given physical quantity with
density $A=A(t, \mbox{\boldmath $x$}, \mbox{\boldmath $v$})$ can
be calculated as $\int A\, f_n\, d^{3n}vd^{3n}x$ we define the
mean value of $A$ in the point $ \mbox{\boldmath $x$}$ through
\begin{equation}
<A(t, \mbox{\boldmath $x$} , \mbox{\boldmath $v$})> \!\!_{v}
=\frac{\int A(t, \mbox{\boldmath $x$} , \mbox{\boldmath $v$}) \,
f_n(t, \mbox{\boldmath $x$} , \mbox{\boldmath $v$} )\, d^{3n}v}
{\int f_n(t, \mbox{\boldmath $x$} , \mbox{\boldmath $v$} )\,
d^{3n}v} \ \ . \label{8}
\end{equation}
This mean value represents the density of $A$ in the space of the
coordinates and permits to write the total value of $A$ as
$\int\rho_{_{\scriptstyle n}}\!\!<\!A\!>\!\!_{v}\,\,d^{3n}x$. We
define the densities in the space of the coordinates for some
quantities used in following. The densities of current, of stress
tensor and of heat flux vector are defined, respectively, by
\begin{eqnarray}
&&\mbox{\boldmath $u$} = \,<\mbox{\boldmath $v$}>\!\!_{v} \ \ ,
\label{9} \\
&&\sigma_{ij}=\mu<(v_i-u_i)(v_j-u_j)> \!\!_{v} \ \ , \label{10} \\
&& h_i =\frac{1}{2}\,\mu<| \mbox{\boldmath $v$}- \mbox{\boldmath
$u$}|^2 (v_i-u_i)>\!\!_{v} \ \ . \label{11}
\end{eqnarray}
The density of energy is given by
\begin{equation}
E=\frac{1}{2}\,\mu<\mbox{\boldmath $v$}^2>\!\!_{v}+ {\cal V }
=\frac{1}{2}\,\mu\, \mbox{\boldmath $u$}^2+\varepsilon+{\cal V} \
\ , \label{12}
\end{equation}
being $\varepsilon>0$ the density of the internal energy.
\begin{equation}
\varepsilon=\frac{1}{2}\,\sigma_{ii}=\frac{1}{2}\,\mu \left
(<\mbox{\boldmath $v$}^2>\!\!_{v}-
 <\mbox{\boldmath $v$}>\!\!_{v}^2 \right ) \ \ . \label{13}
\end{equation}

Using the standard procedure consisting in multiplying Eq.
(\ref{5}) by the three collision invariants $g_k(\mbox{\boldmath
$v$})$ and integrating with respect to $v$, the three following
hydrodynamic equations can be obtained \cite{LI,GL}:
\begin{eqnarray}
&&\frac{\partial \rho_{_{\scriptstyle n}}}{\partial t}
+\frac{\partial }{\partial
\mbox{\boldmath $x$}}\cdot (\,\rho_{_{\scriptstyle n}}
\mbox{\boldmath $u$}\,)=0 \ \ , \label{14} \\
&& \mu\frac{D \mbox{\boldmath $u$} }{D t}=  \mbox{\boldmath ${\cal
F}$}+ \mbox{\boldmath $F$} \ \ , \label{15} \\
&& \frac{\partial}{\partial t}\,(\,\rho_{_{\scriptstyle n}} E\,)
+\frac{\partial }{\partial \mbox{\boldmath $x$}}\cdot
(\,\rho_{_{\scriptstyle n}} \mbox{\boldmath $s$}\,)=0\ \ .
\label{16}
\end{eqnarray}
In the above equations $D/Dt=\partial/\partial t + \mbox{\boldmath
$u$}\cdot\,
\partial/\partial \mbox{\boldmath
$x$}$ is the total time or substantial or Lagrangian derivative,
$\mbox{\boldmath ${\cal F}$}$ is the stress force
\begin{equation}
{\cal F}_i= -\frac{1}{\rho_{_{\scriptstyle n}}}\,\frac{\partial
}{\partial x_j}\, (\,\rho_{_{\scriptstyle n}}\, \sigma_{ij}) \ \ ,
\label{17}
\end{equation}
and $s_i= E u_i + \sigma_{ij}u_j +h_i$ is the energy flux density
vector.

Eq.s (\ref{14})-(\ref{16}) define the most general hydrodynamics
of the particle system which behaves in the physical space as a
fluid. In particular Eq. (\ref{16}) imposes the conservation of
the total energy $H$ of the system
\begin{equation}
H=\int E\, \rho_{_{\scriptstyle n}}\, d^{3n} x \ \ . \label{18}
\end{equation}

After recalling that the external force $\mbox{\boldmath $F$}$ is
conservative, from Eq. (\ref{15}) it follows immediately that the
irrotationality of $\mbox{\boldmath$u$}$ implies the
conservativity of $\mbox{\boldmath ${\cal F}$}$ and viceversa,
namely
\begin{eqnarray}
\mbox{\boldmath $u$}=\frac{1}{\mu}\frac{\partial {\cal S}}
{\partial \mbox{\boldmath $x$}} \label{19} \ \ , \\
\mbox{\boldmath ${\cal F}$} =-\frac{\partial {\cal W
}}{\partial\mbox{\boldmath{$x$}}} \ \ . \label{20}
\end{eqnarray}
The condition (\ref{19}) imposes that the system here considered
is spinless and reduces the vector equation (\ref{15}) into a
scalar one
\begin{equation}
\frac{\partial {\cal S}}{\partial t} +\frac{1}{2\mu}\left
(\frac{\partial {\cal S}}{\partial \mbox{\boldmath $x$}} \right )
^2 + {\cal W} + {\cal V} = 0 \ \ . \label{21}
\end{equation}

{\it Quantum potential:} From the two expressions of the force
$\mbox{\boldmath ${\cal F}$}$ given by Eq.s (\ref{17}) and
(\ref{20}) it follows immediately
\begin{equation}
\frac{\partial {\cal W}}{\partial x_j}= \frac{\partial\,
\sigma_{jk} }{\partial x_k} + \sigma_{jk} \frac{\partial
{\xi}}{\partial x_k}\ \ , \label{22}
\end{equation}
with $\xi=\ln \rho_{_{\scriptstyle n}}$. Note that Eq. (\ref{22})
descends from the requirement that the force $\mbox{\boldmath
${\cal F}$}$ is conservative and can be viewed as a condition
constraining the forms of ${\cal W}$ and $\sigma_{jk}$ which
result to be two functionals of the field $\xi$. The solutions of
the latter equation are couples of ${\cal W}$ and $\sigma_{jk}$
and in principle there can exist more than one solution. Even
though it can appear that Eq. (\ref{22}) contains considerable
degrees of freedom, in the following we will show that the
particular structure of this equation restricts strongly the
number of its solutions.

We suppose now that just as the left hand side, both the first
and the second term in the right hand side in Eq. (\ref{22}) are
curl free and will take the form $\partial(...)/\partial x_j$.
This requirement for the first term can be satisfied in two
different ways. The first and most natural way is related to the
choice $\sigma_{jk}=\sigma \, \delta_{jk}$. Alternatively we can
pose $\sigma_{jk}=\partial a_k/\partial x_j$ and after recalling
the symmetry property $\sigma_{jk}=\sigma_{kj}$ we have that
$a_k=\partial \alpha/\partial x_k$. Then the most general form of
the density of stress tensor is
\begin{equation}
\sigma_{jk}=\sigma \delta_{jk}+\frac{\partial^2 \alpha}{\partial
x_j\partial x_k }\ \ , \label{23}
\end{equation}
being $\sigma$ and $\alpha$ two scalar functionals depending on
$\xi$. After taking into account the identity
\begin{eqnarray}
\!\!\!\!\frac{\partial^2 \alpha }{\partial x_j\partial x_k} \,
\frac{\partial {\xi}}{\partial x_k} =\frac{\partial}{\partial
x_j}\!\left(\frac{\partial{\alpha}}{\partial x_k} \frac{\partial
{\xi}}{\partial x_k} \!\right)-
 \frac{\partial {\alpha}}{\partial x_k}\,
\frac{\partial^2 {\xi}}{\partial x_k \partial x_j} \ , \label{24}
\end{eqnarray}
Eq.(\ref{22}) can be written as
\begin{eqnarray}
\frac{\partial }{\partial x_j}\left({\cal W} - \sigma
-\frac{\partial^2 \alpha }{\partial x_k\partial x_k} -
\frac{\partial{\alpha}}{\partial x_k} \frac{\partial
{\xi}}{\partial x_k} \right) \nonumber \\
= \sigma \frac{\partial {\xi}}{\partial x_j} -
 \frac{\partial {\alpha}}{\partial x_k}\,
\frac{\partial^2 {\xi}}{\partial x_k \partial x_j} \ \ .
\label{25}
\end{eqnarray}
Clearly the two terms in the right hand side of Eq. (\ref{25})
must be curl free. These requirements impose the two following
conditions
\begin{eqnarray}
\frac{\partial{\sigma}}{\partial x_l} \frac{\partial
{\xi}}{\partial x_m}\!&=&\!\frac{\partial{\sigma}}{\partial x_m}
\frac{\partial {\xi}}{\partial x_l} \ \ , \label{26} \\
\frac{\partial^2 \alpha }{\partial x_l\partial
x_k}\,\frac{\partial^2 \xi}{\partial x_k\partial x_m}\!&=&\!
\frac{\partial^2 \alpha }{\partial x_m\partial
x_k}\,\frac{\partial^2 \xi}{\partial x_k\partial x_l} \ \ .
\label{27}
\end{eqnarray}
It is trivial to verify that from Eq. (\ref{26}) it results that
$\sigma=\sigma(\xi)$ is an arbitrary function of $\xi$ while from
Eq. (\ref{27}) one can obtain  that $\alpha=c\xi$ with $c$ a real
arbitrary constant. With these positions Eq. (\ref{25}) becomes
\begin{equation}
\frac{\partial }{\partial x_j}\!\left({\cal W} - \sigma -
\!\!\int\! \!\sigma\, d\xi-c\frac{\partial^2 \xi }{\partial
x_k\partial x_k} - \frac{c}{2}\frac{\partial{\xi}}{\partial x_k}
\frac{\partial {\xi}}{\partial x_k} \right) \! = 0 \ \ .
\label{28}
\end{equation}
From this last equation we can obtain the expression of ${\cal
W}$ while Eq. (\ref{23}) gives the expression of $\sigma_{ij}$.
Finally, we can write the most general solution of Eq. (\ref{22})
under the form
\begin{eqnarray}
&&\!\!\!\!\!\!\!\!\!\! \sigma_{jk}= c \,\,\frac{\partial^{\,2}
\ln \rho_{_{\scriptstyle n}}}{\partial x_j \,\partial x_k} \,+ \,
\delta_{jk} \frac{1}{\rho_{_{\scriptstyle n}}}\!\int\!
\rho_{_{\scriptstyle n}}\frac{d \,{\cal U}(\rho_{_{\scriptstyle
n}})}{d \rho_{_{\scriptstyle n}}}\,\,d\rho_{_{\scriptstyle n}}
 \ \ , \label{29}
\\ && \!\!\!\!\!\!\!\!\!\! {\cal W}=
c\left[ \frac{\partial^2 }{\partial\mbox{\boldmath $x$}^2}\ln
\rho_{_{\scriptstyle n}} +\frac{1}{2} \left( \frac{\partial
}{\partial\mbox{\boldmath $x$}} \ln \rho_{_{\scriptstyle n}}
\right)^{\!\!2} \right ] + {\cal U}(\rho_{_{\scriptstyle n}})\ \
. \label{30}
\end{eqnarray}
The above expressions of $\sigma_{jk}$ and ${\cal W}$ define a
family of solutions of Eq. (\ref{22}) and  represent the
hydrodynamic constitutive equations for the system. Remark that
these equations have been enforced exclusively from the fact that
the internal stress forces are conservative.

In the above constitutive equations the terms containing ${\cal
U}$ describe the standard classical Eulerian fluid. For instance
the nonlinear potential ${\cal U}(\rho_{_{\scriptstyle n}})=a
\rho_{_{\scriptstyle n}}$ appearing in the so called cubic
non-linear Schr\"odinger equation is originated from the stress
tensor $\sigma_{jk}=\frac{1}{2} \,a \,\rho_{_{\scriptstyle
n}}\delta_{jk}$. Analogously the logarithmic non-linearity ${\cal
U}(\rho_{_{\scriptstyle n}})=b \ln \rho_{_{\scriptstyle n}}$ in
the Schr\"odinger equation is originated from a constant stress
tensor $\sigma_{jk}= b \,\delta_{jk}$.

In the following, we pose ${\cal U}=0$ and focalize our attention
to the term of Eqs (\ref{29}) and (\ref{30}), proportional to the
arbitrary constant $c$, which is absent in classical kinetics.
From the definition of the total internal energy, it results that
${\cal H}=\int d^{3n}x \, \rho_{_{\scriptstyle n}} \, \,
\varepsilon > 0$. Starting from Eq. (\ref{29}) and performing an
integration by parts, we obtain ${\cal H}= - c\int d^{3n}x
\,(\partial \rho_{_{\scriptstyle n}} /
\partial \mbox{\boldmath $x$} )^2 /\rho_{_{\scriptstyle n}}$. Then
we pose $c=-\eta^2/4\mu$ in order to put ${\cal H}> 0$. The real
positive constant $\eta$ remains a free parameter of the theory.

We introduce the normalized $n$-particle reduced distribution
function $\varrho_{_{\scriptstyle n}}=\rho_{_{\scriptstyle
n}}/\,N$ being $N=\int \rho_n\,d^{3n} x$,  so that $\int
\varrho_{_{\scriptstyle n}} d^{3n} x=1$. Then we define the mass
$m=N\mu$ the internal potential $W=N{\cal W}$ and set
$\hbar=N\eta$. From (\ref{30}) follows the expression of $W$
\begin{eqnarray}
W= -\frac{\hbar^2}{2m}\,\varrho_{_{\scriptstyle
n}}^{-1/2}\,\frac{\partial^2 \varrho_{_{\scriptstyle
n}}^{1/2}}{\partial\mbox{\boldmath $x$}^2} \ \ . \label{31}
\end{eqnarray}
We recall that the potential $W$ given by Eq. (\ref{31}) is the
quantum potential (Madelug 1926) \cite{MA,BO} and it is
remarkable that here it has been obtained in the framework of an
entirely classical subquantum kinetics. Note that the subquantum
monad structure of the system generates the stress forces and
then the potential $W$. Clearly when this structure is suppressed
the quantum potential vanishes. Furthermore the presence of the
collision integrals $C_i$ in the BBGKY kinetic Eq. (\ref{1})
which take into account the monad interactions are necessary for
the consistency of the theory.

{\it Many particle Schr\"odinger equation:} We define the external
potential $V=N{\cal V}$, and set $S=N{\cal S}$. Now the quantum
fluid is described completely by the two scalar fields
$\varrho_{_{\scriptstyle n}}$ and $S$ whose evolution equations
are (\ref{32}) and (\ref{33}), respectively.
\begin{eqnarray}
&&\!\!\!\!\!\!\!\!\!\!\!\! \frac{\partial \varrho_{_{\scriptstyle
n}}}{\partial t} +\frac{\partial }{\partial \mbox{\boldmath $x$}}
\cdot \left(\varrho_{_{\scriptstyle n}} \frac{1}{m}\frac{\partial
{S}}{\partial \mbox{\boldmath $x$}}\right)=0 \ \ , \label{32} \\
&&\!\!\!\!\!\!\!\!\!\!\!\! \frac{\partial {S}}{\partial t}
+\frac{1}{2m}\left (\frac{\partial {S}}{\partial \mbox{\boldmath
$x$}} \right ) ^2 + W + V = 0 \ \ . \label{33}
\end{eqnarray}
Alternatively, we can describe this fluid by means of the complex
field $\Psi_n=\varrho_{_{\scriptstyle n}}^{1/2}\exp(i{S}/\hbar)$
with $\int |\Psi_n|^2 d^{3n} x=1$, whose evolution equation
\begin{equation}
i\hbar\frac{\partial \Psi_n}{\partial t}= -\frac{\hbar^2}{2m}
\frac{\partial^2\Psi_n}{\partial\mbox{\boldmath $x$}^2}+ V \Psi_n
 \ \ . \label{34}
\end{equation}
can be obtained directly by combining Eq.s (\ref{32}) and
(\ref{33}). This later equation is the $n$-particle Schr\"odinger
equation which, after recalling $V=\sum_iV_i$ with $V_i=N{\cal
V}_i$, assumes the form
\begin{equation}
i\hbar\frac{\partial \Psi_n}{\partial t}=
\sum_{i=1}^n\left(-\frac{\hbar^2}{2m}
\frac{\partial^2\Psi_n}{\partial\mbox{\boldmath $x$}_i^2}+ V_i
\Psi_n \right)
 \ \ . \label{35}
\end{equation}
The total energy $H$ given by Eq. (\ref{18}) results to be the sum
of two terms: $H=H^{(cl)}+ {\cal H}$. The first term
\begin{equation}
H^{(cl)}\!=\!\!\sum_{i=1}^{n}\! \int \! d^{3n} x
\,\varrho_{_{\scriptstyle n}}\frac{1}{2}m \mbox{\boldmath $u$}_i^2
\!+\! \sum_{i=1}^{n}\! \int \! d^{3n} x \,\varrho_{_{\scriptstyle
n}}{ V}_i(\mbox{\boldmath $x$}) \ , \label{36}
\end{equation}
corresponds to the energy of $n$ interacting classical particles
while the second term ${\cal H}=\int \varepsilon
\rho_{_{\scriptstyle n}} \,\,d^{3n} x$ is originated by the
internal structure of the $n$ bodies and is given by
\begin{eqnarray}
{\cal H}=\frac{\hbar^2}{8m}\int  d^{3n} x
\,\,\frac{1}{\varrho_{_{\scriptstyle n}}} \,
 \left( \frac{\partial {\varrho_{_{\scriptstyle n}}}}{\partial
\mbox{\boldmath $x$}} \right )^2=\frac{\hbar^2}{8m}\,\,I \ \ .
\label{37}
\end{eqnarray}
Eq. (\ref{37}) hints on the existence of a tight link between
quantum mechanics and Fisher information theory. It is remarkable
that ${\cal H}$ results to be proportional to the Fisher
information measure $I$ (Fisher 1922) which describes an
important feature of quantum mechanics \cite{RH,RE,FR,FR2}. Once
again, within the scenario of the subquantum structure of quantum
particle, another basic physical concept, namely $I$, emerges
naturally. In the present context $I$ has a very transparent
meaning being simply a measure of the internal energy of quantum
system. The fact that ${\cal H}$ is originated from the internal
structure of the $n$-bodies becomes more transparent if we
observe that ${\cal H}$ can be written also as
\begin{eqnarray}
{\cal H}=\sum_{i=1}^{n} \frac{\hbar^2}{8m}\int  d^{3} x_i
\,\,\frac{1}{\varrho_{_{\scriptstyle n}}} \,
 \left( \frac{\partial {\varrho_{_{\scriptstyle n}}}}{\partial
\mbox{\boldmath $x$}_i} \right
)^2=\frac{\hbar^2}{8m}\sum_{i=1}^{n}I_i \ \ , \label{38}
\end{eqnarray}
from which it results $I=\sum_iI_i$ and then ${\cal
H}=\sum_i{\cal H}_i$. An analogous property holds also for the
quantum potential $W=\sum_i W_i$ being
\begin{eqnarray}
W_i= -\frac{\hbar^2}{2m}\,\varrho_{_{\scriptstyle
n}}^{-1/2}\,\frac{\partial^2 \varrho_{_{\scriptstyle
n}}^{1/2}}{\partial\mbox{\boldmath $x$}_i^2} \ \ . \label{39}
\end{eqnarray}
Of course starting from $H=H^{(cl)}+ {\cal H}$ and taking into
account Eq.s (\ref{35}) and (\ref{36}) one obtains immediately the
well known expression of the Hamiltonian of the $n$-particle
quantum mechanics
\begin{eqnarray}
H=&&\!\!\!\!\!\!\int d^{3n} x { \left (\frac{\hbar ^2}{2m}\left |
\frac{\partial \Psi_n} {\partial \mbox{\boldmath $x$}} \right |
^2 + V \, |\Psi_n|^2 \right )} = \sum_{i=1}^n H_i \ \ . \label{40}
\end{eqnarray}

In the present picture the mass of the $n$-body quantum system is
$M=nm$. After recalling that $m=N\mu$ follows immediately that
the system is composed by ${\cal N}=nN$ monads.

{\it Conclusions:} The approach used here to obtain the
Schr\"odinger equation for a system of $n$ quantum particles
starting from the BBGKY hierarchy describing a statistical system
of ${\cal N}=nN$ subquantum monads, suggests that quantum
mechanics can be viewed as a hidden variables theory. This
approach can be used easily also to construct the non linear
quantum mechanics [it is enough to consider ${\cal U}\neq 0$ in
Eqs (\ref{29}),(\ref{30})] and then for study the non linear
effects originated for instance from the particle bosonic or
fermionic nature \cite{GKA,GKE}.


\begin{references}

\bibitem{GK}
G. Kaniadakis, Physica A {\bf 307}, 172 (2002);

quant-ph/0112049.
\bibitem{GK2}
G. Kaniadakis, Found. Phys. Lett. {\bf 16}(2), 99 (2003);

quant-ph/0209033.
\bibitem{CG}
R. Czopnik and P. Garbaczewski, Phys. Lett. A {\bf 299}, 447
(2002).
\bibitem{BA}
R. Balescu, {\it Statistical Dynamics: Matter out of
Equilibrium}, Imperial College Press, London (2000).
\bibitem{LI}
R. L. Liboff, {\it Kinetics Theory}, Prentice-Hall International
Editors, N.J. (1990).
\bibitem{GL}
R. T. Glassey, {\it The Cauchy problem in Kinetic Theory}, SIAM,
Philadelphia, PA (1996).
\bibitem{MA}
E. Madelung, Z. Phys. {\bf 40}, 322 (1926).
\bibitem{BO}
D. Bohm, Phys. Rev. {\bf 85}, 166 (1952).
\bibitem{RH}
J. Rehacek and Z. Hradil, Phys. Rev. Lett. {\bf 88}, 130401
(2002).
\bibitem{RE}
M. Reginatto, Phys. Rev. A {\bf 58}, 1775 (1998).
\bibitem{FR}
B.R. Frieden and B.H. Soffer, Phys. Rev. {\bf 52}, 2274 (1995).
\bibitem{FR2}
B.R. Frieden, {\it Physics from Fisher Information: A
Unification}, Cambridge University Press, Cambridge (1999).
\bibitem{GKA}
G. Kaniadakis, Phys. Rev. A {\bf 55}, 941 (1997).
\bibitem{GKE}
G. Kaniadakis and A.M. Scarfone, Phys. Rev. E {\bf 64}, 026106
(2001).


\end{references}
\end{document}